\documentclass[ALICE,manyauthors]{cernphprep}

\usepackage[comma,square,numbers,sort&compress]{natbib}
\usepackage{hyperref}
\usepackage{url}
\usepackage{lineno}
\usepackage{epsfig} 
\usepackage{epstopdf} 
\usepackage{comment}
\usepackage{amsmath}
\usepackage{amsthm}
\usepackage{amsfonts}
\usepackage{amssymb}
\usepackage{xspace}
\usepackage{datetime}
\usepackage{mathtools}
\usepackage{mathrsfs}

\usepackage[english,colorinlistoftodos]{todonotes}
\presetkeys{todonotes}{inline, color=blue!30}{}

\newcommand{\jpsi}{\ensuremath{\text{J}/\psi}\xspace}

\newcommand{\pp}{\ensuremath{{\rm pp}}\xspace}

\newcommand{\s}{\ensuremath{\sqrt{s}}\xspace}

\newcommand{\kT}{\ensuremath{k_{\text{T}}}}
\newcommand{\pt}{\ensuremath{p_{\text{T}}}}
\newcommand{\pT}{\ensuremath{p_{\text{T}}}}
\newcommand{\mee}{\ensuremath{m_{\rm ee}}} 

\newcommand{\der}{\ensuremath{{\rm d}}}

\newcommand{\ccBar}{\ensuremath{\text{c}\overline{\text{c}}}\xspace}
\newcommand{\bbBar}{\ensuremath{\text{b}\overline{\text{b}}}\xspace}
\newcommand{\dEdx}{\ensuremath{\text{d}E/\text{d}x}\xspace}

\newcommand{\ele}{\ensuremath{\text{e}^{-}}\xspace}
\newcommand{\pos}{\ensuremath{\text{e}^{+}}\xspace}

\newcommand{\AccEff}{\ensuremath{\langle A\times \epsilon \rangle}}

\newcommand{\LumiInt}{\ensuremath{L_{\rm int}}\xspace}
\newcommand{\BR}{\ensuremath{\text{BR}(\jpsi \to \pos \ele)}}

\newcommand{\chisqdof}{\ensuremath{\chi^{2}/{N_{\rm dof}}}}

\newcommand{\sigmae}{\ensuremath{\sigma_{\rm e}}}

\newcommand{\MeanPt}{\ensuremath{\langle p_{\text{T}} \rangle}}
\newcommand{\MeanPtSq}{\ensuremath{\langle p_{\text{T}}^{2} \rangle}}
\newcommand{\GeVc}{\ensuremath{{\rm GeV}/c}\xspace}

\newcommand{\GeVcsq}{\ensuremath{{\rm GeV}/c^2}}
\newcommand{\MeVcsq}{\ensuremath{{\rm MeV}/c^2}}
\newcommand{\diele}{\ensuremath{{\rm e}^{+}{\rm e}^{-}}}

\newcommand{\dSigmadPtdy}{\ensuremath{\der^{2} \sigma / \der \pT \der y}\xspace}
\newcommand{\dSigmady}{\ensuremath{\der \sigma / \der y}\xspace}
\newcommand{\fB}{\ensuremath{f_{\mathrm{B}}}}

\def\Version#1{\def\version{#1}}
\Version{0.0}

\begin{document}%

\begin{titlepage}
\PHyear{2019}
\PHnumber{014}      
\PHdate{12 May}  
%

\title{Inclusive $\jpsi$ production at mid-rapidity in pp collisions at $\sqrt{\pmb{s}}$ = 5.02 TeV}
\ShortTitle{Inclusive $\jpsi$ production in $\pp$ collisions at $\s=5.02$~TeV}   

\Collaboration{ALICE Collaboration\thanks{See Appendix~\ref{app:collab} for the list of collaboration members}}
\ShortAuthor{ALICE Collaboration} 

\begin{abstract}
Inclusive $\jpsi$ production is studied in minimum-bias proton-proton collisions at a centre-of-mass energy of $\s = 5.02$~TeV by ALICE at the CERN LHC. The measurement is performed at mid-rapidity ($|y| < 0.9$) in the dielectron decay channel down to zero transverse momentum $\pT$, using a data sample corresponding to an integrated luminosity of $\LumiInt = 19.4 \pm 0.4$~nb$^{-1}$. The measured $\pt$-integrated inclusive $\jpsi$ production cross section is $\dSigmady = 5.64 \pm 0.22 \text{(stat.)} \pm 0.33 \text{(syst.)} \pm 0.12 \text{(lumi.)}$~$\mu$b. The $\pt$-differential cross section $\dSigmadPtdy$ is measured in the $\pt$ range $0-10~\GeVc$ and compared with state-of-the-art QCD calculations. The $\jpsi$ $\MeanPt$ and $\MeanPtSq$ are extracted and compared with results obtained at other collision energies. 
\end{abstract}
\end{titlepage}
\setcounter{page}{2}

\section{Introduction}
\label{sec:Intro}

At LHC energies, charmonium is mainly produced from gluon--gluon scatterings producing $\ccBar$ pairs~\cite{Mangano:1997ri} which form a bound state. While the hard gluon--gluon scattering can be described within perturbative Quantum Chromodynamics (QCD), the hadronisation of the $\ccBar$ pair into charmonium is essentially non-perturbative and cannot be yet calculated from the QCD Lagrangian. There are several phenomenological approaches for the description of charmonium production: the Colour Evaporation Model (CEM)~\cite{CEM1, CEM2}, the Colour Singlet Model (CSM)~\cite{CSM1} and the Non-Relativistic QCD model (NRQCD)~\cite{NRQCD1} which differ mainly in the way the charmonium states are formed in the hadronisation process. In the CEM model, the production rate of a given charmonium state is proportional to the production cross section of $\ccBar$ pairs integrated between $m_{\ccBar}$ and twice the mass of the lightest D-meson, where $m_{\ccBar}$ is twice the mass of the charm quark, or, according to a recent conjecture ~\cite{Ma:2016exq, Cheung:2018tvq}, the mass of the bound state itself. In the CSM model, the pre-resonant $\ccBar$ state is assumed to be directly produced colourless and with the same quantum numbers as the final-state charmonium. The NRQCD model includes all possible colour and quantum number states for the pre-resonant $\ccBar$ pair, with each configuration having a probability to transform into a given bound state, described by a set of universal long-distance matrix elements determined from global fits to experimental data. Detailed reviews of the state-of-the-art calculations for charmonium production can be found in Refs.~\cite{Brambilla:2011epjc, Brambilla:2014jmp,Andronic:2015wma}.

$\jpsi$ production is a probe of the hot and dense medium created in ultrarelativistic heavy-ion collisions~\cite{Kluberg:2009wc,BraunMunzinger:2009ih}. Moreover, it is also sensitive to nuclear effects not related to the creation of deconfined matter, called cold-nuclear-matter effects, such as modification of the parton distribution functions~\cite{Eskola:2016oht,Kovarik:2015cma}. In order to gauge both the hot and cold medium effects, precise knowledge of the $\jpsi$ production rates in the absence of a nucleus in the initial state is of paramount importance. The $\jpsi$ measurement in $\pp$ collisions constitutes a baseline for the quantification of nuclear effects in both nucleus--nucleus and proton--nucleus collisions.

In this paper, results for the transverse momentum ($\pt$) dependence of the inclusive $\jpsi$ production cross section at mid-rapidity ($|y| < 0.9$) in $\pp$ collisions at the centre-of-mass energy $\s~=5.02$~TeV are presented. The inclusive cross section contains a prompt contribution, which includes directly produced $\jpsi$ as well as the feed-down from the prompt decay of heavier charmonium states (mainly $\psi(2S)$ and $\chi_c$), and a non-prompt contribution from the weak decay of beauty hadrons.

The $\jpsi$ production is measured in the dielectron decay channel using the ALICE central barrel detectors. The $\pt$-differential cross section is measured for $\pt<10~\GeVc$ supplementing the existing mid-rapidity measurements at high $\pt$ by ATLAS~\cite{Aaboud:2017cif} and CMS~\cite{Sirunyan:2017mzd} down to zero $\pt$. Thus, ALICE can measure the $\pt$-integrated inclusive $\jpsi$ production cross section, the mean transverse momentum $\MeanPt$ and the second moment of the transverse momentum $\MeanPtSq$. Similar measurements in $\pp$ collisions were performed by ALICE at $\s=2.76$~TeV~\cite{Abelev:2012kr} and at $\s=7$~TeV~\cite{Aamodt:2011gj} at mid- and forward rapidity ($2.5<y<4.0$), and at $\s=5.02$, $8$ and $13$~TeV at forward-rapidity~\cite{Adam:2016rdg,ALICE8TeV,Acharya:2017hjh}. Prompt $\jpsi$ production cross sections were measured at $\s = 7$~TeV by ALICE~\cite{Abelev:2012gx} and LHCb~\cite{Aaij:2012asz} and at $\s = 8$ and $13$~TeV by LHCb~\cite{Aaij:2013yaa,Aaij:2015rla}.

The paper is organised as follows: the ALICE apparatus and the data sample are described in Section~\ref{sec:Exp_Data}, the data analysis is detailed in Section~\ref{sec:Data_analysis} and the results are discussed in Section~\ref{sec:CrossSec} in comparison with other measurements and theoretical calculations. Conclusions are given in Section~\ref{sec:Conclusions}.

\section{Apparatus and data sample}
\label{sec:Exp_Data}

The central barrel of the ALICE detector~\cite{Aamodt:2008zz,Abelev:2014ffa} allows the reconstruction of $\jpsi$ in the $\diele$ decay channel at mid-rapidity. The entire setup is placed in a solenoidal magnetic field of $B=0.5$~T oriented along the beam direction. 

In this analysis, the Inner Tracking System (ITS) \cite{ITS} and the Time Projection Chamber (TPC)~\cite{2010NIMPA.622..316A} are used for tracking whereas the TPC provides the electron identification. The ITS is subdivided into six cylindrically-shaped layers of silicon detectors around the beam pipe with radii from $3.9$ to $43.0$~cm. The two innermost layers form the high granularity Silicon Pixel Detector (SPD), the two intermediate layers the Silicon Drift Detector (SDD), and the outermost layers the Silicon Strip Detector (SSD). The ITS provides precise tracking close to the interaction point and collision vertex position determination. The TPC is a large drift detector with a cylindrical geometry which extends radially between $85<r<250$~cm and longitudinally between $-250<z<250$~cm, where $z=0$ and $r=0$ correspond to the nominal interaction point. It is the main tracking device, with a full azimuthal acceptance for tracks in the pseudorapidity range $|\eta|<0.9$. Additionally, the TPC can be used for the particle identification of charged particles via the measurement of the specific ionisation energy loss $\dEdx$ in the TPC gas.

The minimum bias (MB) trigger is provided by the V0 detector which consists of two forward scintillator arrays~\cite{Abbas:2013taa} placed on both sides of the nominal interaction point at $z=-90$ and $+340$~cm covering the $\eta$ range $-3.7 < \eta < -1.7$ and $2.8 < \eta < 5.1$. The trigger signal consists of a coincident signal on both sides and is fully efficient in inelastic collisions containing a $\jpsi$. 

For this analysis, the data recorded by ALICE in the 2017 LHC pp run at a centre-of-mass collision energy of $\s=5.02$~TeV are used. A total of 987 million MB events are used in this analysis corresponding to an integrated luminosity of $\LumiInt=19.4 \pm 0.4$~nb$^{-1}$. The integrated luminosity is obtained following a procedure~\cite{lumi5TeV} which employs the Van der Meer technique~\cite{VanDerMeer}.

\section{Analysis, corrections, and systematic uncertainties}
\label{sec:Data_analysis}
%
%
\subsection{Event and track selection}
\label{sec:Selection}
The $\jpsi$ candidates are searched in the dielectron decay channel, with the electron tracks being reconstructed in the ITS and the TPC. The events fulfill the MB trigger condition and have the collision vertex within the longitudinal interval $|z_{\rm vtx}|<10$~cm to ensure uniform detector acceptance. Beam-gas events are rejected using offline timing cuts with the V0 detector. The probability for collision pile-up was $\leq$1\% during the entire data taking period and these events are rejected using a vertex finding algorithm based on SPD tracklets~\cite{Abelev:2014ffa}. 

%
%
Electron candidates are required to have a minimum transverse momentum of $1~\GeVc$ and a pseudorapidity in the range of $|\eta | < 0.9$. Due to the short decay time of the $\jpsi$ and its decay mothers, if any, the daughter electrons are reconstructed as primary particles \cite{ALICEprimaryParticle}. The candidate daughter tracks are required to have a maximum distance-of-closest-approach to the reconstructed collision vertex of $0.2$~cm in the radial direction and $0.4$~cm along the beam-axis direction. Monte Carlo (MC) simulations are used to verify that this requirement does not reject electrons from the decays of non-prompt $\jpsi$. Tracks which originate from long-lived weak decays of charged particles (e.g. $\pi^{\pm} \rightarrow \mu^{\pm}\nu$ or $K^{\pm} \rightarrow \mu^{\pm}\nu$) are rejected from the analysis.
A hit in at least one of the two SPD layers is required for both electron candidates to improve the tracking resolution and reduce the number of electrons from photon conversions. Electron candidates are required to have at least $70$ out of a maximum of $159$ attached clusters and the track fit $\chi^2/N_{\mathrm{dof}}<2$  in the TPC.

Electron candidates are selected such that their specific ionisation energy loss $\dEdx$ in the TPC lies within the interval $[-2, +3]~\sigmae$ relative to the expectation for electrons with same momentum as the candidate, where $\sigmae$ is the specific energy-loss resolution for electrons in the TPC. Similarly, to further reject contamination, particles compatible within $3\sigma$ with being a proton or a pion, according to the measured $\dEdx$, are rejected. 

The dominant source of background electrons is photon conversions. Electrons from conversions in the material at large radii (typically beyond the SPD layers) are removed using the requirement on the SPD hits described above. Electrons from conversions occuring in the beam pipe or in the SPD material can pass the primary track selection criteria. Therefore, further rejection of this background is done by employing a method which relies on a second set of electrons selected with looser criteria. Electrons from the first (primary) set are paired with those of the second set. For pairs with an invariant mass below the threshold of $50~\MeVcsq$, the corresponding electron from the primary set is excluded from further analysis. The looser selection criteria of the second set are optimized in a data driven way such that the signal to background ratio is improved, but the loss of signal with respect to not applying this procedure remains negligible.

%
%
\subsection{Signal extraction}
\label{sec:Signal_extraction}
\begin{figure}[!t]
  \centering
  \includegraphics[scale = 0.6]{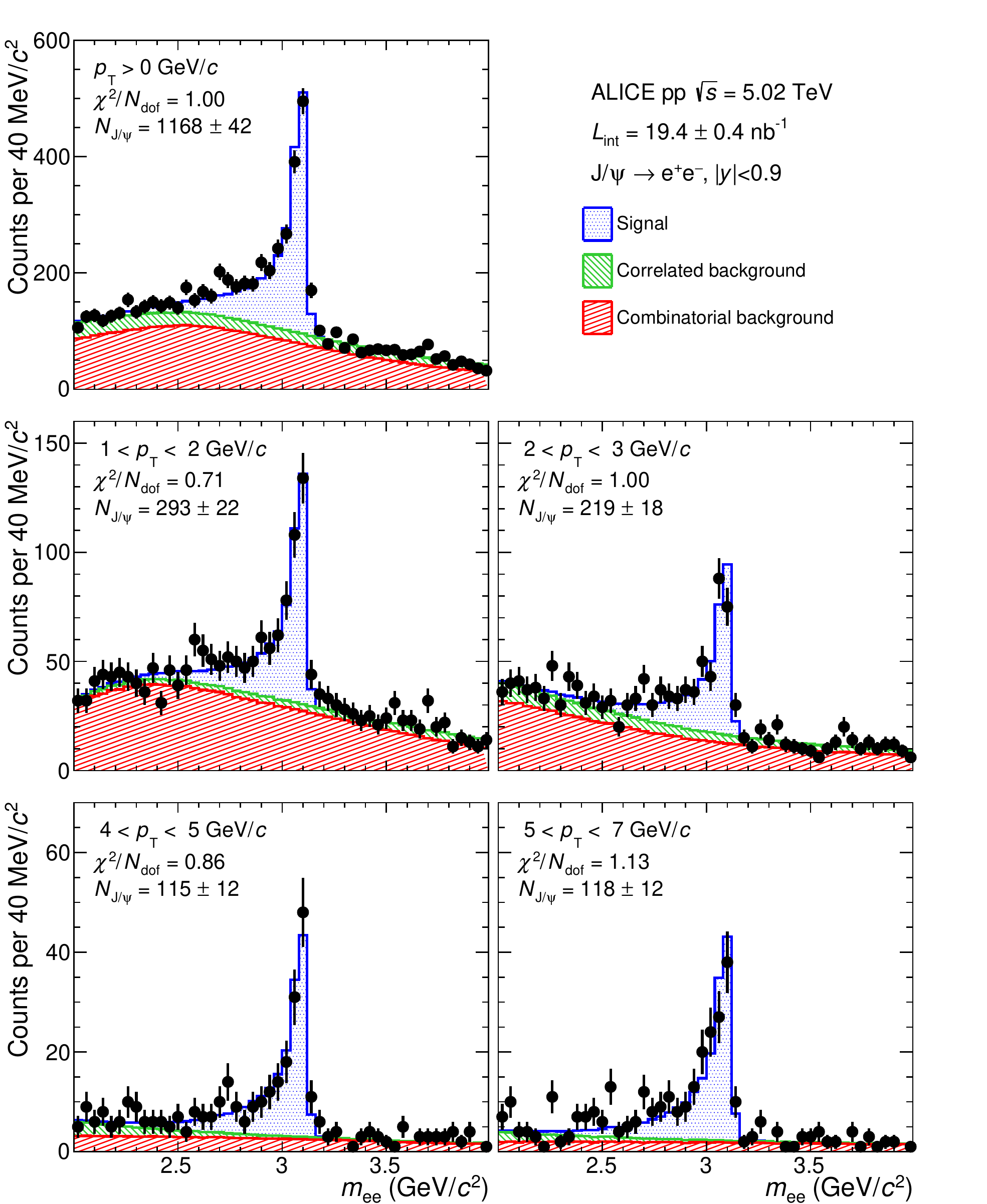}
  \caption{(Colour online) Same-event opposite-sign dielectron invariant mass distributions for several $\pt$-intervals with signal (blue), correlated background (green), and combinatorial background (red) components.}\label{fig:InvMass}
\end{figure}
The $\jpsi$ signal is extracted from the invariant mass distribution of all opposite-sign pairs obtained by making all possible combinations with the electrons and positrons selected with the criteria described above. Examples of invariant mass distributions of opposite-sign (OS) electron pairs are shown in Fig.~\ref{fig:InvMass} for the $\pt$-integrated case and for a few selected $\pt$ intervals. These distributions contain contributions from the $\jpsi$ signal and the combinatorial and correlated backgrounds. For the combinatorial background, kinematic correlations do not play a significant role and this component can be modelled using a mixed event technique, while the correlated background in the $\jpsi$ mass region originates mainly from semi-leptonic decays of correlated open heavy-flavour hadrons~\cite{Acharya:2018ohw}. The signal component corresponds to the electron pairs from $\jpsi$ decays and has an asymmetric shape due to the radiative component and to the energy lost by the electrons in the detector material via brehmsstrahlung.

In order to obtain the raw number of $\jpsi$ counts, a two-step procedure is employed. First, the combinatorial background is obtained using a mixed event (ME) technique and scaled such that the invariant mass distribution of like-sign (LS) pairs from ME matches the same-event LS distribution in the invariant mass range $1.2<\mee<5.0$~$\GeVcsq$. Second, the combinatorial background is subtracted and the remaining distribution is fit with a two-component function, an exponential (or a second order polynomial) for the correlated background and the MC template of the $\jpsi$ signal shape. This strategy provides a good fit quality for all the $\pt$ intervals, as indicated by the $\chi^2/N_{\mathrm{dof}}$ values shown in the panels of Fig.~\ref{fig:InvMass}. The number of $\jpsi$ candidates is obtained by counting the bin entries in the mass interval $2.92<\mee<3.16~\GeVcsq$ after subtracting all background components. The background-subtracted signal distribution is also fit with a Crystal Ball function~\cite{PhysRevD.34.711} and the $\pt$-integrated dielectron mass resolution at the $\jpsi$ peak region obtained from the Gaussian core of the function is found to be $23$~$\MeVcsq$.

Alternative fit strategies for the same event OS invariant mass distribution are considered. A first strategy is to make a template fit to the total OS invariant mass distribution, where the mixed event LS background is used as the template for the combinatorial background with the normalisation used as a free parameter, while the correlated background and signal components are defined similarly as for the standard method. A second alternative is to fit the OS mass distribution with the MC template for the signal component, while for the sum of the combinatorial and correlated background an ad-hoc function is used (ratio of small order polynomials). These alternative methods produce compatible results.

\subsection{Corrections}
\label{sec:Corrections}
In order to correct the observed $\jpsi$ signal for detector effects and the selection procedure, MC events are generated by adding a single $\jpsi$ meson to a simulated MB pp collision. PYTHIA 6.4~\cite{Sjostrand:2006za} is used to simulate the MB events and the non-prompt $\jpsi$ component, while the prompt component is produced uniformly distributed in rapidity with a $\pT$ spectrum based on a phenomenological interpolation of measurements at RHIC, CDF, and the LHC \cite{Bossu}. The $\jpsi$ decays, including the radiative component, are handled by PHOTOS~\cite{Golonka:2005pn}. The transport through the ALICE detector material is handled by GEANT3~\cite{Brun:1994aa} with tracks being reconstructed from the simulated hits using the same algorithm as for the real data. The $\pt$-integrated acceptance times efficiency $\AccEff$ is $9.9$\%, varying between $8.1$\% and $13$\% as a function of $\pt$, and is the product of the acceptance factor, the reconstruction efficiency including the track quality cuts, the electron identification cuts and the fraction of the signal within the mass counting interval of $2.92 < \mee  < 3.16~\GeVcsq$. Due to the variation of $\AccEff$ with the $\jpsi$ transverse momentum in the considered $\pT$ intervals, the calculated correction factors have a mild dependence on the shape of the $\pT$ distribution and the fraction of non-prompt $\jpsi$, $\fB$, used in the simulation. In order to correct for this effect, the corrected $\jpsi$ cross-section, obtained initially using efficiencies weighted with the $\jpsi$ spectrum from simulations, is used to reweight the acceptance times efficiency factor and obtain an updated cross-section. This procedure can be iteratively employed  until the variation of the $\jpsi$ cross section between two iterations is smaller than a desired precision. 
Already after the first iteration, the inclusive $\jpsi$ $\pt$-integrated cross section varied by less than $1$\%, while for the $\pt$-differential cross section the changes were even smaller, so the procedure is stopped after one iteration.
Due to the fact that in our analysis there is a $1$--$2$\% difference in acceptance times efficiency between prompt and non-prompt $\jpsi$ and that the $\fB$ value in simulation is larger with respect to existing measurements at Tevatron~\cite{Acosta:2004yw} and LHC~\cite{Khachatryan:2010yr,Abelev:2012gx,Aad:2015duc} energies, the acceptance times efficiency factors are reweighted to account for these differences. The largest impact from this correction is observed at high $\pt$, where the difference between simulation and existing $\fB$ measurements is largest, and shifts the cross section upwards by $0.3$\%.  

%
%
The differential cross section in a rapidity interval $\Delta y$ and transverse momentum interval $\Delta \pt$ is calculated as
\begin{equation}
  \frac{\der^2 \sigma_{\jpsi}}{\der y \der \pt} = \frac{N_{\jpsi} (\Delta y , \Delta \pt)}{\BR \cdot \AccEff (\Delta y , \Delta \pt) \cdot \Delta y \cdot \Delta \pt \cdot \LumiInt},
\end{equation}
where $N_{\text{J}/\psi}$ is the number of reconstructed $\jpsi$ candidates, $\BR$ is the branching ratio of the $\jpsi$ mesons decaying into dielectrons~\cite{PDG2018}, and $\LumiInt$ is the integrated luminosity of the data sample. 

%
%
\subsection{Systematic uncertainties}
\label{sec:Systematics}

The sources of systematic uncertainties are related to the ITS-TPC tracking, electron identification, signal extraction procedure,  the $\jpsi$ input kinematic distributions used in the MC production, the integrated luminosity determination, and the branching ratio of the dielectron decay channel. A summary of all the systematic uncertainties is provided in Table~\ref{tab:Syst}.

\begin{table}[!b]
  \centering
 \caption{Summary of the contributions to the systematic uncertainty (in percentage) for the inclusive $\pt$-integrated cross section $\dSigmady$ and in the different $\pT$ intervals.
  All sources of systematic uncertainty are considered to be highly correlated over $\pT$, except for the background fit which is considered fully uncorrelated.}\label{tab:Syst}
  
  \begin{tabular}{ l | c c c c c c c c }
    \multicolumn{1}{c|}{Source} & \multicolumn{8}{c}{$\pT$~($\GeVc$)} \\
    \cline{2-9}
                   & $\pt>0$ & 0$-$1 & 1$-$2 & 2$-$3 & 3$-$4 & 4$-$5 & 5$-$7 & 7$-$10    \\
    \hline
    \hline
    Tracking       & 5.3     & 4.8 & 5.1 & 5.4 & 5.5 & 5.2 & 5.6 & 5.7     \\
    PID            & 0.43    & 0.21 & 0.42 & 0.17 & 0.04 & 0.40 & 1.1 & 2.5   \\
    Signal shape   & 1.9     & 1.8 & 1.8 & 1.8 & 1.8 & 1.8 & 2.9 & 2.9   \\
    Background fit & 0.21    & 1.0 & 0.30 & 0.36 & 0.18 & 0.27 & 0.21 & 0.21    \\
    MC input       & 1.4     & 0.24 & 0.35 & 0.10 & 0.15 & 0.11 & 0.15 & 0.69   \\
    \multicolumn{1}{l|}{Luminosity} & \multicolumn{8}{c}{2.1}  \\
    \multicolumn{1}{l|}{Branching ratio} & \multicolumn{8}{c}{0.53}  \\
    \hline
    Total uncorrelated syst.  & 0.21     & 1.0 & 0.30 & 0.36 & 0.18 & 0.27 & 0.21 & 0.21     \\
    Total correlated syst.    & 5.8     & 5.1 & 5.5 & 5.7 & 5.8 & 5.6 & 6.4 & 6.9     \\
    \multicolumn{1}{l|}{Global syst.} & \multicolumn{8}{c}{2.2}  \\
    \hline
    Total          & 6.2     & 5.6 & 5.9 & 6.1 & 6.2 & 6.1 & 6.8 & 7.3  \\
  \end{tabular}
  
\end{table}

The dominant source of systematic uncertainty is related to the ITS-TPC tracking and has two components, one related to the ITS-TPC matching efficiency and the other to the track quality requirements. The component due to the ITS-TPC matching efficiency is the largest and is determined by comparing the probability to match the TPC tracks to hits in the ITS in both data and simulation \cite{ALICEDmesonPP7}. 

After propagation to the $\jpsi$ candidate pairs, this uncertainty is found to vary between $4.3$\% at low $\pt$ up to $5.4$\% at high $\pt$. The uncertainty due to the track quality requirements amounts to approximately $2$\% in all $\pt$ intervals and was obtained by varying the selection criteria and computing the RMS of the cross-section distribution obtained after these variations. This tracking uncertainty is considered to be correlated over $\pt$.

The systematic uncertainty due to the electron identification is estimated by comparing the response of the TPC electron identification of a clean sample of electrons from tagged photon conversions in data to true electrons from the MC simulation. Half the difference between the selection efficiency in data and simulation is taken as the systematic uncertainty on the single electron PID efficiency and propagated to that on the $\jpsi$ selection using a toy MC simulating $\jpsi$ decays in the dielectron channel. 

The uncertainty due to the TPC PID ranges between $0.1$\% at intermediate and $2.5$\% at high $\pt$ and is considered to be correlated over $\pt$.

The uncertainty on the signal extraction procedure has contributions from the choice of the $\jpsi$ invariant mass shape and from the fit procedure used to describe the correlated background. It is estimated by varying the mass interval used for the signal counting and the mass range used for the fitting. The value of the uncertainty is determined as the RMS of the distribution of cross sections obtained from the cases which give statistically significant variations, similar to the procedure described by Barlow~\cite{Barlow}. The uncertainty on the $\jpsi$ signal shape ranges between $1.8$\% in the low- and $2.9$\% in the high-$\pt$ intervals and is treated to be correlated over $\pt$. The uncertainty due to the background fitting is $1$\% in the lowest $\pt$-interval and less than $0.5$\% otherwise. It is considered as uncorrelated.

The uncertainty from the $\jpsi$ $\pt$-distribution which is used to compute the corrections is related to the precision of the fit to the measured $\jpsi$ spectrum which is used in the iterative procedure described in Section~\ref{sec:Corrections}. The fit parameters are varied randomly within their allowed fit uncertainty taking into account their correlation matrix. The resulting uncertainty amounts to $1.4$\% for the $\pt$-integrated cross section and less than $1$\% in each of the considered $\pt$-intervals. 

The systematic uncertainty on the integrated beam luminosity is described in detail in Ref.~\cite{lumi5TeV} and amounts to $2.1$\%. This uncertainty is taken as a global uncertainty for the $\pt$-integrated and the $\pt$-differential cross sections.

The uncertainty on the branching ratio $\BR = (5.97 \pm 0.03)$\%~\cite{PDG2018} is treated as fully correlated between all bins.

\section{Results}
\label{sec:CrossSec}

The inclusive $\jpsi$ cross section in $\pp$ collisions at $\s=5.02$~TeV measured at mid-rapidity in the interval $|y|<0.9$ is
\begin{equation*}
\der \sigma_{\jpsi} / \der y = 5.64 \pm 0.22 \text{(stat.)} \pm 0.33 \text{(syst.)} \pm 0.12 \text{(lumi.)}~\mu {\rm b}.
\end{equation*}
The systematic uncertainty contains all the sources described in Section~\ref{sec:Data_analysis} added in quadrature, assuming that the $\jpsi$ is produced unpolarised. Although the existing measurements in $\pp$ collisions at LHC energies indicate a null or only a small polarisation~\cite{ALICEpolariz7TeV, Acharya:2018uww, LHCbPolarization7TeV}, there are no polarisation measurements for $\jpsi$ at low $\pt$ and mid-rapidity at LHC energies. In order to estimate the impact on the measured inclusive $\jpsi$ cross section, the acceptance and efficiency factors are reweighted to take into account various polarisation scenarios. In the extreme cases of a fully transverse ($\lambda=+1$) or a fully longitudinal ($\lambda=-1$) polarisation in the helicity frame, the $\pt$-integrated cross section would increase by $15$\% or decrease by $24$\%, respectively.

\begin{figure}[!t]
  \centering
  \includegraphics[scale = 0.39]{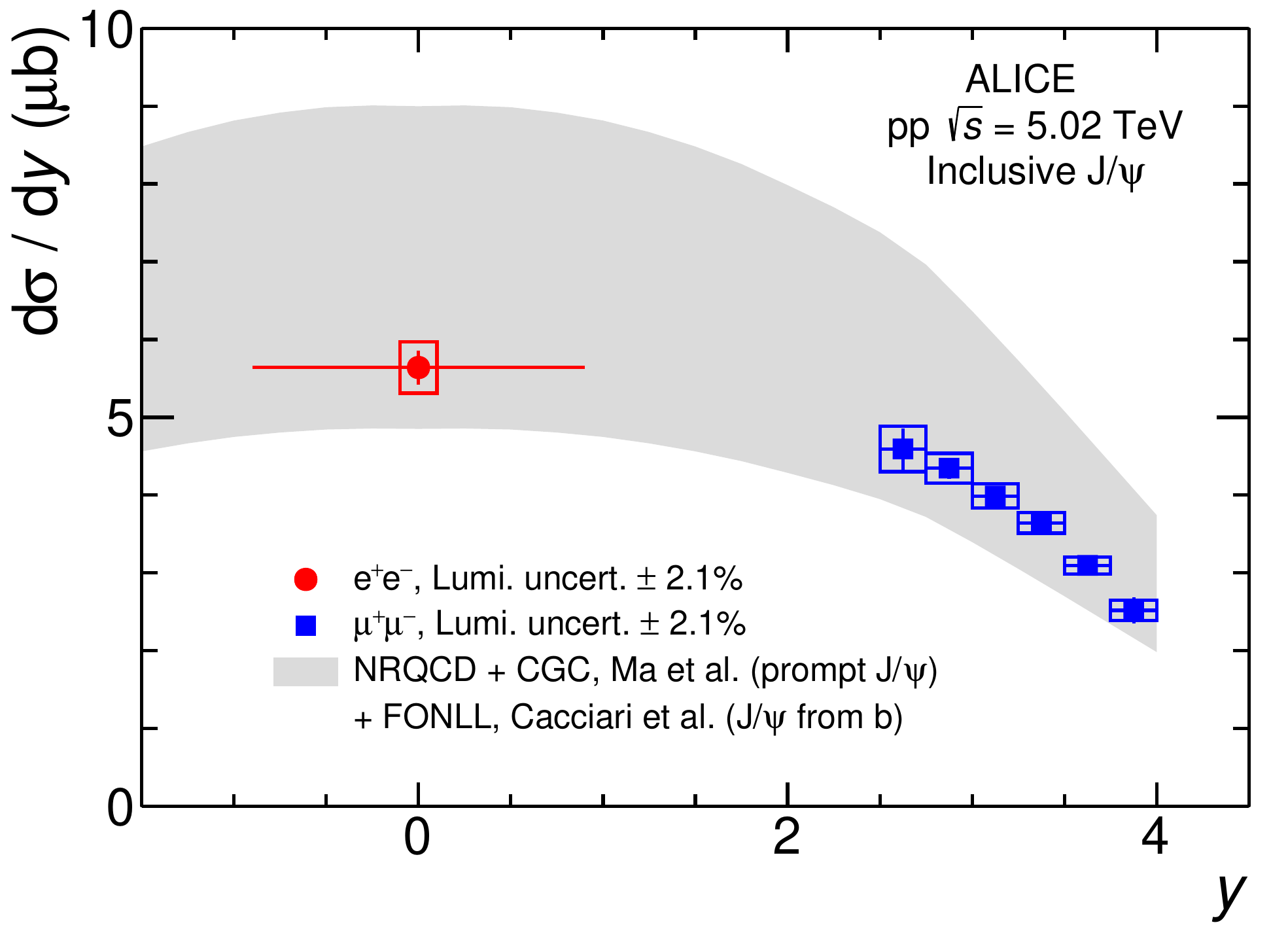}
  \includegraphics[scale = 0.39]{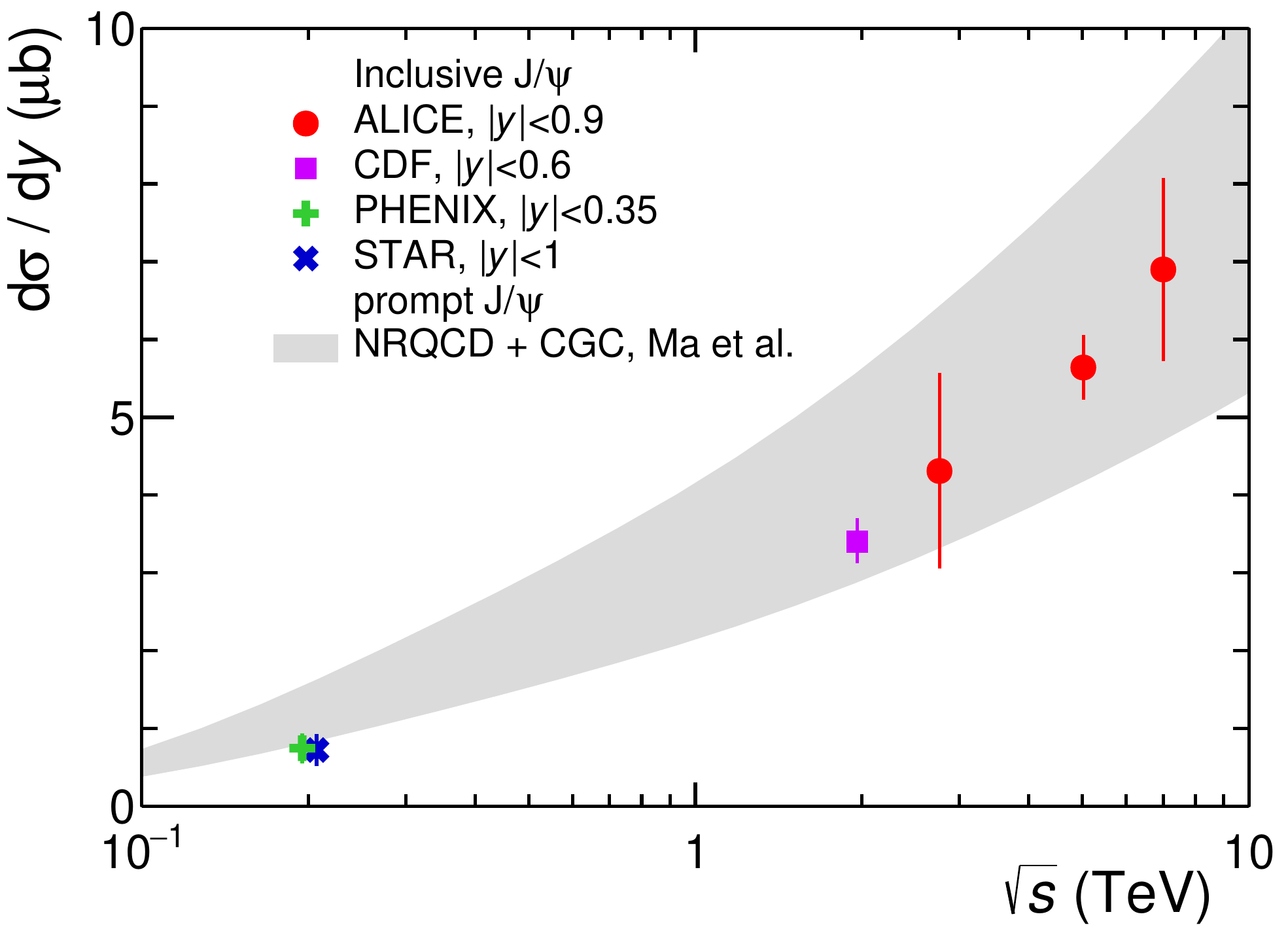}
  \caption{Left: Inclusive $\jpsi$ cross section as a function of rapidity compared to the ALICE results at forward rapidity~\cite{Acharya:2017hjh} and to calculations from~\cite{NRQCD:Venugopalan} to which a non-prompt component is added as computed in~\cite{FONLL}. Right: Inclusive $\jpsi$ cross section at mid-rapidity~\cite{PHENIX200GeV,Adam:2018jmp,Acosta:2004yw,Abelev:2012kr,Aamodt:2011gj} as a function of collision energy compared to the calculations from ~\cite{NRQCD:Venugopalan}. The data points from PHENIX and STAR, both at $\s=0.2$~TeV, are slightly shifted for improved visibility.}\label{fig:CrossSec}
\end{figure}

In the left panel of Fig.~\ref{fig:CrossSec}, the inclusive $\pt$-integrated cross section $\dSigmady$ is compared with the ALICE measurements at forward rapidity in the dimuon channel~\cite{Acharya:2017hjh}. The systematic uncertainties are represented as boxes and the statistical uncertainties are shown by vertical error bars. The reported $\jpsi$ cross sections are inclusive and contain both the prompt and non-prompt components. The rapidity-dependent cross section is compared with results for prompt $\jpsi$ from Leading Order (LO) NRQCD calculations coupled to a Colour Glass Condensate (CGC) description of the gluon distributions in the proton from Ma and Venugopalan~\cite{NRQCD:Venugopalan}. This model includes a soft-gluon resummation which allows the calculation of the \jpsi cross section down to zero \pt. The Long Distance Matrix Elements (LDME) used are obtained by fitting the prompt component of high-$\pT$ $\jpsi$ at Tevatron~\cite{Chao:2012iv}. Feed-down from higher mass charmonia, $\psi$(2S) and $\chi_c$, are considered. The non-prompt component is calculated with Fixed-Order Next-to-Leading Logarithm (FONLL)~\cite{FONLL} from beauty quarks with a $\jpsi$ in the final state. The prompt component from Ref.~\cite{NRQCD:Venugopalan} and the non-prompt component from Ref.~\cite{FONLL} are then added together in order to generate the inclusive $\jpsi$ cross section shown in the left panel of Fig.~\ref{fig:CrossSec}. The uncertainties of the prompt and non-prompt component are assumed to be uncorrelated when calculating the error band of the sum. The non-prompt contribution to the inclusive cross section is of the order of $10$--$20$\% in the considered low-$\pT$ regime. The relatively large uncertainty band of the model is mainly due to variations of the charm-quark mass, and the renormalisation and factorisation scales. Assuming that the rapidity dependence in the calculation is not affected by the change of these scales, the rapidity dependence of the $\jpsi$ cross section is well reproduced in the model. The overall normalisation of the calculation has very large uncertainties and these data represent a strong constrain to the model assumptions.

The energy dependence of the $\jpsi$ cross section in $\pp$ collisions at mid-rapidity is shown in the right panel of Fig.~\ref{fig:CrossSec}. The results are compared with the PHENIX~\cite{PHENIX200GeV} and STAR~\cite{Adam:2018jmp} measurements at $\s=0.2$~TeV, the CDF measurement at $\s=1.96$~TeV~\cite{Acosta:2004yw}, and previous ALICE measurements at $\s=2.76$~\cite{Abelev:2012kr} and $7$~TeV~\cite{Aamodt:2011gj}, where statistical and systematic uncertainties are added in quadrature. A steady increase, approximately logarithmic in $\s$, of $\dSigmady$ at mid-rapidity is observed. The data are compared with the calculated prompt $\jpsi$ cross section from Ref.~\cite{NRQCD:Venugopalan}. Since the non-prompt component is known to be of the order of $10$\% of the inclusive cross section, the qualitative comparison to the data is not affected. As in the case of the rapidity dependence discussed above, the calculations are compatible with the logarithmic trend seen in the data, while the absolute normalisation has large uncertainties. 

\begin{figure}[!b]
  \centering
  \includegraphics[scale = 0.39]{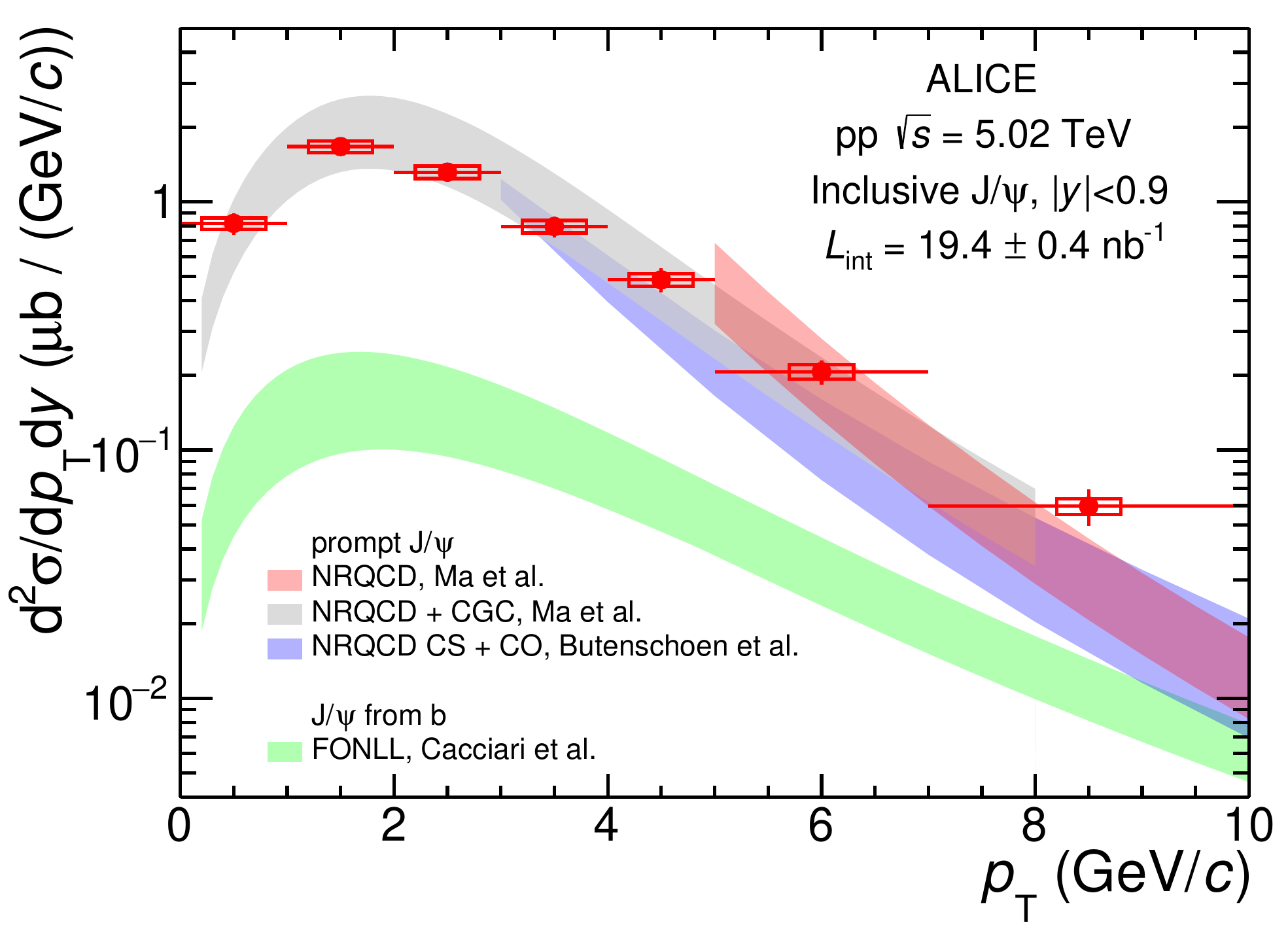}
  \includegraphics[scale = 0.39]{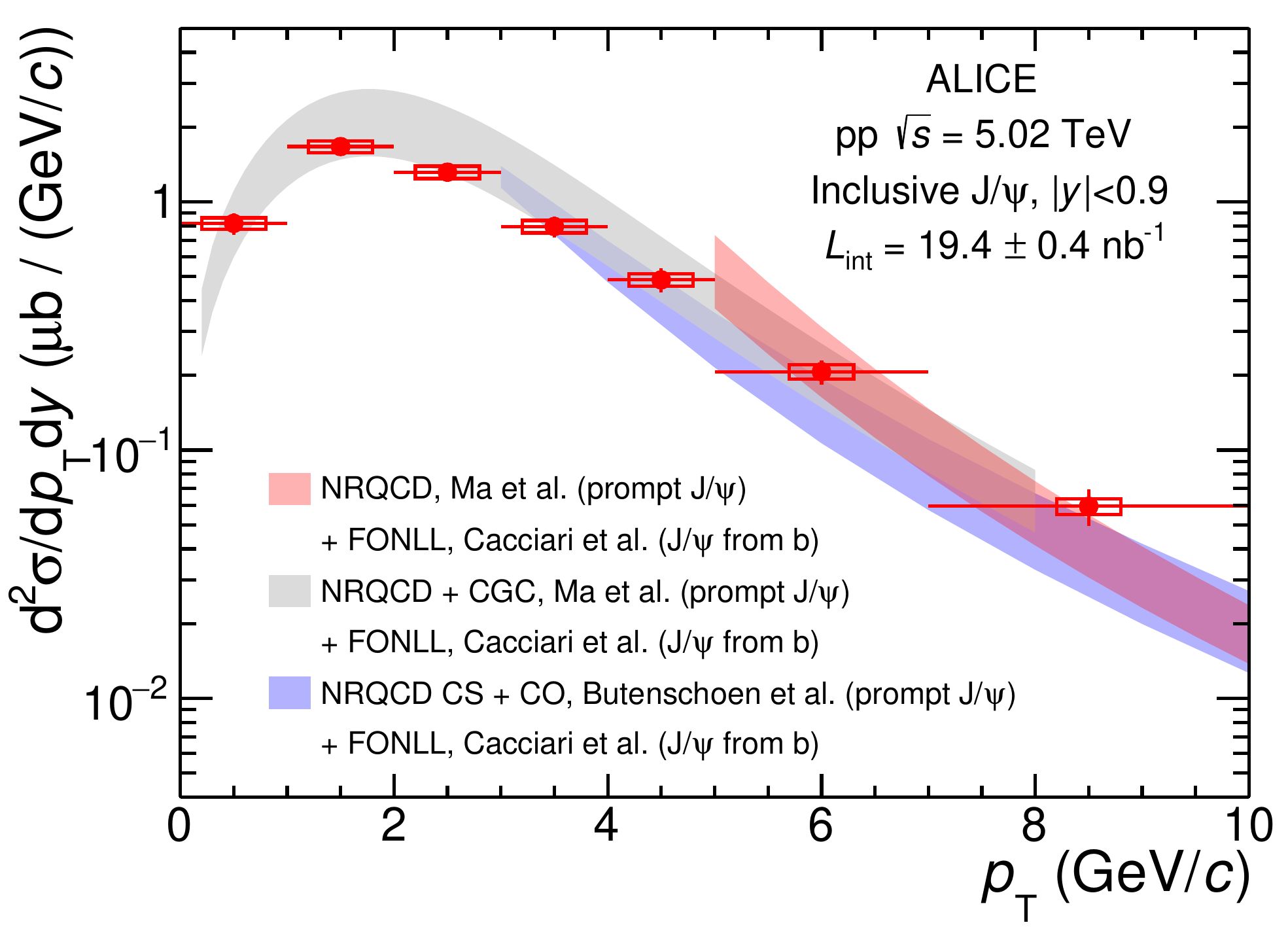}
  \caption{$\pT$-differential inclusive $\jpsi$ cross section compared with prompt $\jpsi$ calculations from NLO NRQCD~\cite{NRQCD:Ma,NRQCD:Kniehl} and LO NRQCD+CGC~\cite{NRQCD:Venugopalan} and non-prompt $\jpsi$ calculations from FONLL~\cite{FONLL}. The calculations for the prompt and non-prompt components are shown separately in the left panel while in the right panel the FONLL calculation is added to the prompt $\jpsi$ calculations.}\label{fig:Theory}
\end{figure}

In the left panel of Fig.~\ref{fig:Theory}, the $\pT$-differential cross section $\dSigmadPtdy$ is compared to three calculations of the prompt $\jpsi$ cross section: two NLO NRQCD calculations from Ma et al.~\cite{NRQCD:Ma} and Butenschoen et al.~\cite{NRQCD:Kniehl}, and the above-mentioned calculations usind leading order NRQCD and CGC~\cite{NRQCD:Venugopalan}. The non-prompt component obtained using FONLL~\cite{FONLL} is shown separately. In the right panel of Fig.~\ref{fig:Theory}, the non-prompt FONLL calculation is added to each of the three prompt calculations and compared with results of the present analysis. Within the model uncertainties, the NRQCD+CGC model provides a good description of the trend over the covered $\pt$ interval, with the lower part of the band being favoured by the data. Although employing a very similar approach for the small distance coefficients, the two NLO NRQCD calculations use quite different LDME values, extracted by fitting charmonium cross-sections measured at Tevatron and HERA with different low $\pt$ cut-offs. This limits the range of validity to $\pt>3$ and $\pt>5$~\GeVc for the cross sections obtained in Ref.~\cite{NRQCD:Kniehl} and Ref.~\cite{NRQCD:Ma}, respectively. In addition, the calculations from Ref.~\cite{NRQCD:Kniehl} predict a strong transversal $\jpsi$ polarization, which is in contradiction to the recent ALICE measurement at forward-rapidity in pp collisions at $\s=8$~TeV~\cite{Acharya:2018uww} which favours zero or a small amount of polarisation. The cross sections from Ref.~\cite{NRQCD:Kniehl} do not include feed-down contributions from higher mass charmonia. Both predictions are in agreement to the data considering the uncertainties, however, the above mentioned differences in model assumptions together with the large scale uncertainties prevent drawing firm conclusions. There are recent alternative works, not at the presented energy, using an improved CEM model~\cite{Ma:2016exq, Cheung:2018tvq} or NRQCD in the $\kT$ factorisation approach~\cite{Baranov:2011ib,Baranov:2015laa,Baranov:2015yea} that could further help interpret our data.

\begin{figure}[!t]
  \centering
  \includegraphics[scale = 0.5]{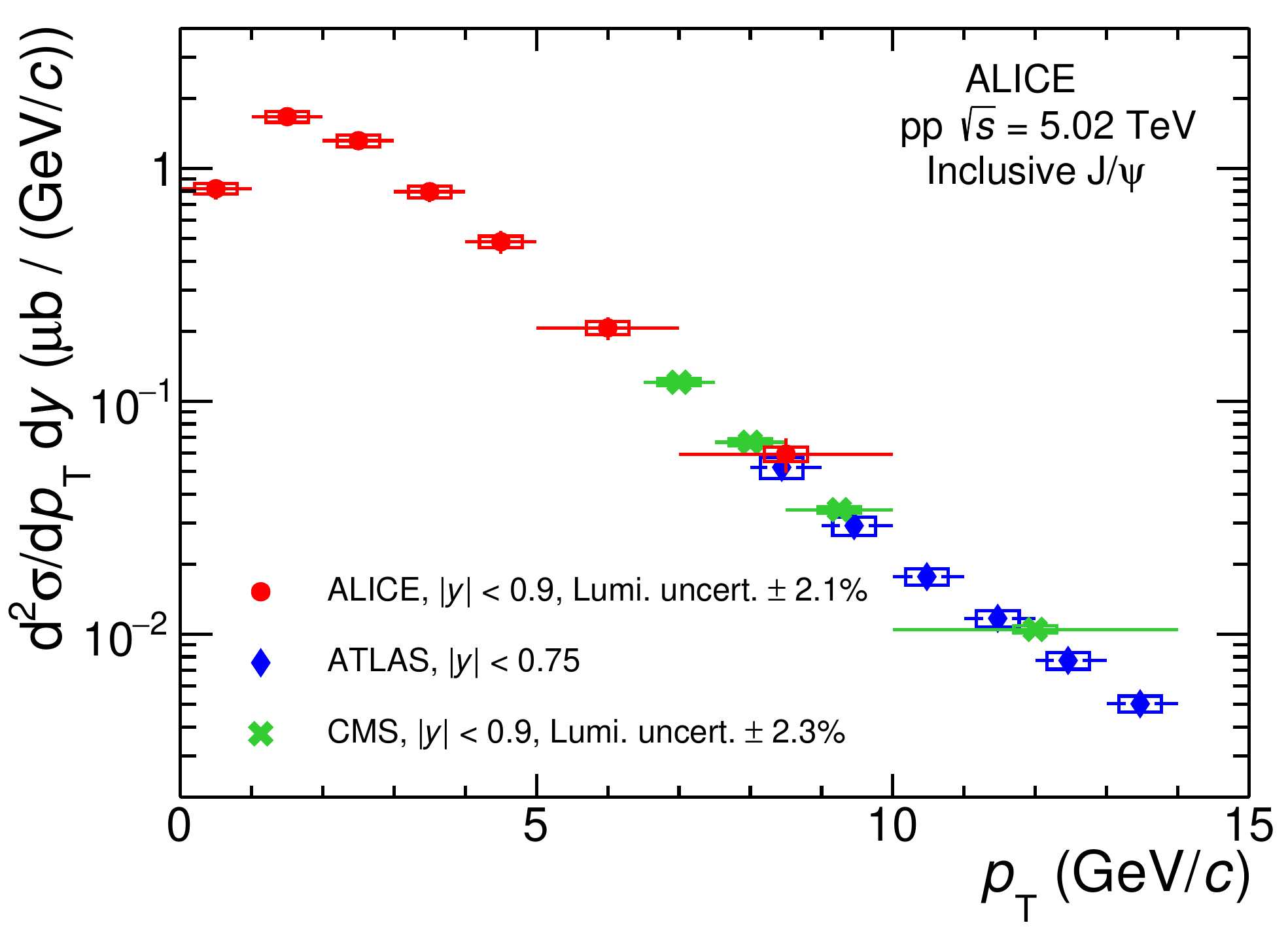}
  \caption{$\pT$-differential inclusive $\jpsi$ cross section compared with ATLAS~\cite{Aaboud:2017cif} and CMS~\cite{Sirunyan:2017mzd} results at mid-rapidity. Luminosity uncertainties are indicated in the legend, except in the case of ATLAS for which these are included in the boxes.}
  \label{fig:CrossSec_Pt}
\end{figure}

In Fig.~\ref{fig:CrossSec_Pt}, the $\pT$-differential cross section $\dSigmadPtdy$ is compared with the high-$\pt$ measurements from ATLAS~\cite{Aaboud:2017cif} and CMS~\cite{Sirunyan:2017mzd} at mid-rapidity and same collision energy. It should be noted that the ATLAS and CMS measurements extend to higher $\pT$ but are truncated to a region which is relevant for the comparison to ALICE. The ATLAS and CMS measurements of the prompt and non-prompt contributions were summed in order to obtain the inclusive cross section needed to compare with our measurement. Good agreement is observed between the results in the overlapping $\pt$ region.

The energy dependence of the $\pT$-differential $\jpsi$ cross section can be studied by using its moments, the average transverse momentum $\MeanPt$ and the squared average transverse momentum $\MeanPtSq$. In this analysis, the inclusive $\jpsi$ $\MeanPt$ and $\MeanPtSq$ are obtained by fitting the measured spectrum with a power law function of the form
\begin{equation}
   f(\pT) = C \times \frac{\pT}{\left\{ 1 + \left(\pT/p_0 \right)^2 \right\}^n },
   \label{eq:PowerLaw}
\end{equation}
where $C$, $p_0$, and $n$ are free fit parameters. In the interval $\pt<10~\GeVc$, the first two moments of the fitted function are
\begin{eqnarray*}
  \MeanPt &=& 2.66 \pm 0.06 \text{(stat.)} \pm 0.01 \text{(syst.)} ~\GeVc,\\
  \MeanPtSq &=& 10.2 \pm 0.5 \text{(stat.)} \pm 0.1 \text{(syst.)} ~{\rm GeV}^{2}/c^2.
\end{eqnarray*}
The systematic uncertainty is obtained by fitting the measured $\jpsi$ spectrum only with the systematic uncertainty of the $\pT$-differential cross section. The statistical uncertainty on the $\MeanPt$ and $\MeanPtSq$ takes into account the correlation matrix of the parameters from the fit procedure. A cross check of these results is performed considering a fit to the dielectron $\MeanPt$ and $\MeanPtSq$ distribution as a function of the invariant mass. A polynomial fit function is used to parameterise the background $\MeanPt$ and $\MeanPtSq$ as a function of invariant mass, and the signal-over-background ratio obtained from the signal extraction procedure as discussed in Section~\ref{sec:Signal_extraction}. The values obtained with this cross check are found to be compatible with the ones obtained from the spectrum fit.
\begin{figure}[!t]
  \centering
  \includegraphics[scale = 0.5]{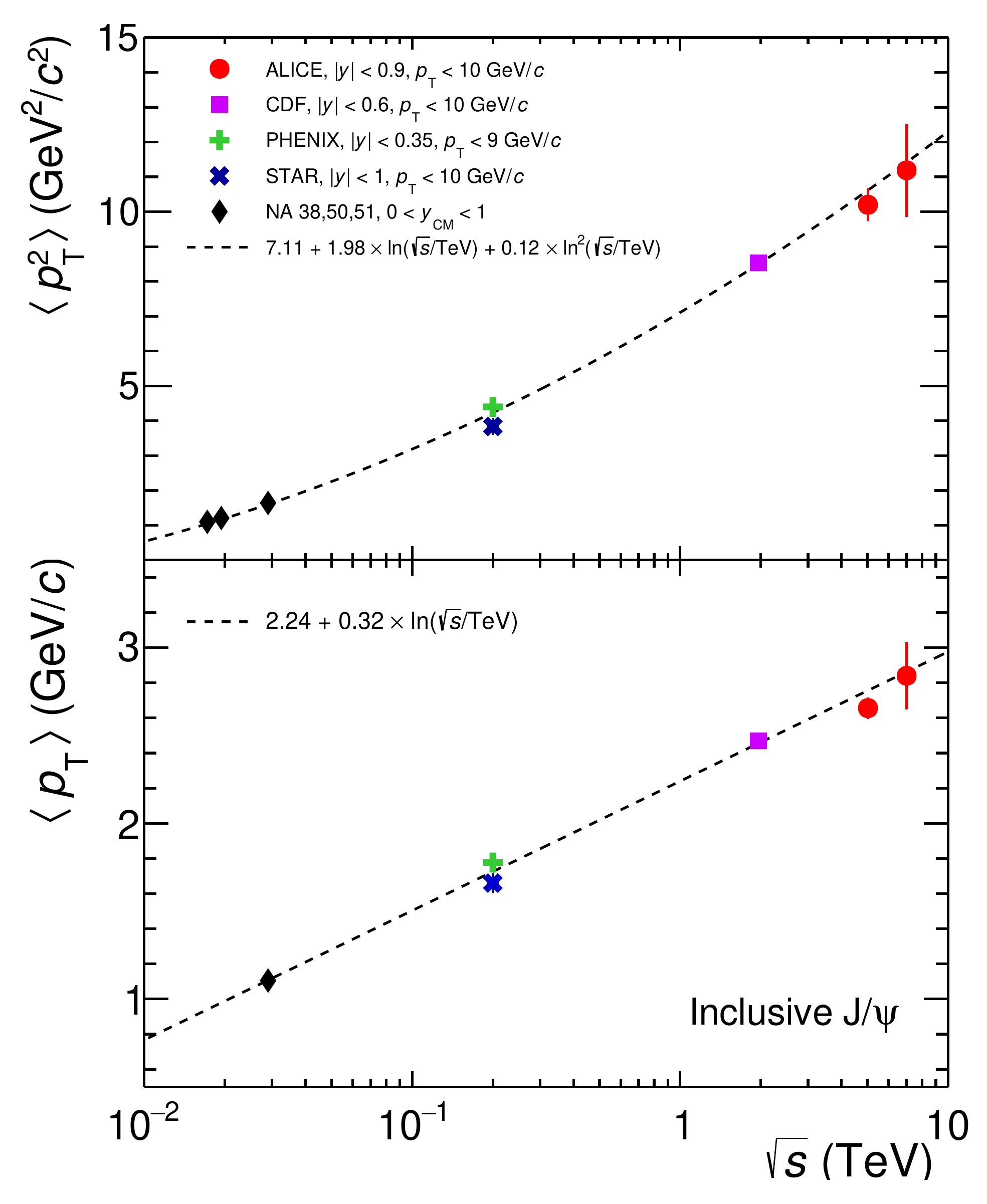}
  \caption{The $\MeanPt$ (lower panel) and $\MeanPtSq$ (upper panel) of inclusive $\jpsi$ at mid-rapidity as a function of the collision energy. These results are compared to previous results from ALICE at LHC~\cite{Aamodt:2011gj, Abelev:2012gx}, Tevatron~\cite{Acosta:2004yw}, RHIC~\cite{PHENIX200GeV,Adam:2018jmp} and SPS~\cite{SPSmeanpt}.}\label{fig:MeanPtvsSqrtS}
\end{figure}

The energy dependences of the $\MeanPt$ and $\MeanPtSq$ moments are shown in Fig.~\ref{fig:MeanPtvsSqrtS}. A steady increase with energy is observed for both $\MeanPt$ and $\MeanPtSq$ over a wide collision energy range which includes results from SPS~\cite{SPSmeanpt}, RHIC~\cite{PHENIX200GeV,Adam:2018jmp}, Tevatron~\cite{Acosta:2004yw}, and LHC~\cite{Aamodt:2011gj, Abelev:2012gx}. Statistical and systematic uncertainties are added in quadrature. This behaviour is a consequence of the opening of the phase space with increasing collision energy, i.e. for a fixed Bjorken-$x$ the momentum-exchange $Q^2$ grows with increasing collision energy leading to a hardening of the $\jpsi$ $\pt$ spectrum. Also, the faster increase with energy of the $\bbBar$ cross section compared to the $\ccBar$ cross section leads to a growth of the non-prompt $\jpsi$ fraction, which further hardens the $\jpsi$ $\pT$ spectrum. In order to quantify the energy dependence of the $\jpsi$ $\MeanPt$ and $\MeanPtSq$, we performed similar fits to those used in Ref.~\cite{PHENIX200GeV, ALBAJAR1990261}, where linear or quadratic functions of the logarithm of the centre-of-mass collision energy were used. These simple parametrisations describe the existing measurements over nearly three orders of magnitude in collision energy, with values of the $\chisqdof$ of 1.7 and 0.98 for the $\jpsi$ $\MeanPt$ and $\MeanPtSq$, respectively. 

\section{Conclusions}
\label{sec:Conclusions}
The inclusive $\jpsi$ production cross section in proton--proton collisions at $\s = 5.02$~TeV in the rapidity range $|y| < 0.9$ is measured down to zero $\pt$ using the dielectron decay channel. The measurement is performed using a minimum-bias data sample corresponding to an integrated luminosity of $\LumiInt = 19.4 \pm 0.4$~nb$^{-1}$ and yields a $\pt$-integrated cross section of $\dSigmady = 5.64 \pm 0.22 \text{(stat.)} \pm 0.33 \text{(syst.)} \pm 0.12 \text{(lumi.)}$~$\mu$b.

Comparisons of the inclusive $\pt$-integrated and $\pt$-differential cross section of three NRQCD calculations for prompt $\jpsi$ summed with a non-prompt $\jpsi$ component calculated with FONLL are compatible with the data if the large scale uncertainties are considered as uncorrelated over rapidity, collision energy or $\pt$. A more refined approach, which would consider correlations between model parameters will allow to differentiate between the different theoretical approaches. 

A good agreement to the complementary ATLAS and CMS measurements at the same collision energy is observed in the overlapping $\pt$ interval. The energy dependence of the $\MeanPt$ and $\MeanPtSq$ indicate a hardening of the $\pt$-differential cross section with increasing collision energy. This is well described by a linear and squared logarithmic increase of $\MeanPt$ and $\MeanPtSq$ with $\s$, respectively.

\newenvironment{acknowledgement}{\relax}{\relax}
\begin{acknowledgement}
\section*{Acknowledgements}

The ALICE Collaboration would like to thank all its engineers and technicians for their invaluable contributions to the construction of the experiment and the CERN accelerator teams for the outstanding performance of the LHC complex.
The ALICE Collaboration gratefully acknowledges the resources and support provided by all Grid centres and the Worldwide LHC Computing Grid (WLCG) collaboration.
The ALICE Collaboration acknowledges the following funding agencies for their support in building and running the ALICE detector:
A. I. Alikhanyan National Science Laboratory (Yerevan Physics Institute) Foundation (ANSL), State Committee of Science and World Federation of Scientists (WFS), Armenia;
Austrian Academy of Sciences, Austrian Science Fund (FWF): [M 2467-N36] and Nationalstiftung f\"{u}r Forschung, Technologie und Entwicklung, Austria;
Ministry of Communications and High Technologies, National Nuclear Research Center, Azerbaijan;
Conselho Nacional de Desenvolvimento Cient\'{\i}fico e Tecnol\'{o}gico (CNPq), Universidade Federal do Rio Grande do Sul (UFRGS), Financiadora de Estudos e Projetos (Finep) and Funda\c{c}\~{a}o de Amparo \`{a} Pesquisa do Estado de S\~{a}o Paulo (FAPESP), Brazil;
Ministry of Science \& Technology of China (MSTC), National Natural Science Foundation of China (NSFC) and Ministry of Education of China (MOEC) , China;
Croatian Science Foundation and Ministry of Science and Education, Croatia;
Centro de Aplicaciones Tecnol\'{o}gicas y Desarrollo Nuclear (CEADEN), Cubaenerg\'{\i}a, Cuba;
Ministry of Education, Youth and Sports of the Czech Republic, Czech Republic;
The Danish Council for Independent Research | Natural Sciences, the Carlsberg Foundation and Danish National Research Foundation (DNRF), Denmark;
Helsinki Institute of Physics (HIP), Finland;
Commissariat \`{a} l'Energie Atomique (CEA), Institut National de Physique Nucl\'{e}aire et de Physique des Particules (IN2P3) and Centre National de la Recherche Scientifique (CNRS) and R\'{e}gion des  Pays de la Loire, France;
Bundesministerium f\"{u}r Bildung und Forschung (BMBF) and GSI Helmholtzzentrum f\"{u}r Schwerionenforschung GmbH, Germany;
General Secretariat for Research and Technology, Ministry of Education, Research and Religions, Greece;
National Research, Development and Innovation Office, Hungary;
Department of Atomic Energy Government of India (DAE), Department of Science and Technology, Government of India (DST), University Grants Commission, Government of India (UGC) and Council of Scientific and Industrial Research (CSIR), India;
Indonesian Institute of Science, Indonesia;
Centro Fermi - Museo Storico della Fisica e Centro Studi e Ricerche Enrico Fermi and Istituto Nazionale di Fisica Nucleare (INFN), Italy;
Institute for Innovative Science and Technology , Nagasaki Institute of Applied Science (IIST), Japan Society for the Promotion of Science (JSPS) KAKENHI and Japanese Ministry of Education, Culture, Sports, Science and Technology (MEXT), Japan;
Consejo Nacional de Ciencia (CONACYT) y Tecnolog\'{i}a, through Fondo de Cooperaci\'{o}n Internacional en Ciencia y Tecnolog\'{i}a (FONCICYT) and Direcci\'{o}n General de Asuntos del Personal Academico (DGAPA), Mexico;
Nederlandse Organisatie voor Wetenschappelijk Onderzoek (NWO), Netherlands;
The Research Council of Norway, Norway;
Commission on Science and Technology for Sustainable Development in the South (COMSATS), Pakistan;
Pontificia Universidad Cat\'{o}lica del Per\'{u}, Peru;
Ministry of Science and Higher Education and National Science Centre, Poland;
Korea Institute of Science and Technology Information and National Research Foundation of Korea (NRF), Republic of Korea;
Ministry of Education and Scientific Research, Institute of Atomic Physics and Ministry of Research and Innovation and Institute of Atomic Physics, Romania;
Joint Institute for Nuclear Research (JINR), Ministry of Education and Science of the Russian Federation, National Research Centre Kurchatov Institute, Russian Science Foundation and Russian Foundation for Basic Research, Russia;
Ministry of Education, Science, Research and Sport of the Slovak Republic, Slovakia;
National Research Foundation of South Africa, South Africa;
Swedish Research Council (VR) and Knut \& Alice Wallenberg Foundation (KAW), Sweden;
European Organization for Nuclear Research, Switzerland;
National Science and Technology Development Agency (NSDTA), Suranaree University of Technology (SUT) and Office of the Higher Education Commission under NRU project of Thailand, Thailand;
Turkish Atomic Energy Agency (TAEK), Turkey;
National Academy of  Sciences of Ukraine, Ukraine;
Science and Technology Facilities Council (STFC), United Kingdom;
National Science Foundation of the United States of America (NSF) and United States Department of Energy, Office of Nuclear Physics (DOE NP), United States of America.
\end{acknowledgement}
\newpage

\bibliographystyle{utphys}   
\bibliography{alicepreprint_jpsi}

\newpage
\appendix
\section{The ALICE Collaboration}
\label{app:collab}

\begingroup
\small
\begin{flushleft}
S.~Acharya\Irefn{org141}\And 
D.~Adamov\'{a}\Irefn{org93}\And 
S.P.~Adhya\Irefn{org141}\And 
A.~Adler\Irefn{org74}\And 
J.~Adolfsson\Irefn{org80}\And 
M.M.~Aggarwal\Irefn{org98}\And 
G.~Aglieri Rinella\Irefn{org34}\And 
M.~Agnello\Irefn{org31}\And 
N.~Agrawal\Irefn{org48}\textsuperscript{,}\Irefn{org10}\And 
Z.~Ahammed\Irefn{org141}
% ...
\And
\renewcommand\labelenumi{\textsuperscript{\theenumi}~}

\section*{Affiliation notes}
\renewcommand\theenumi{\roman{enumi}}
\begin{Authlist}
\item \Adef{org*}Deceased
\item \Adef{orgI}Dipartimento DET del Politecnico di Torino, Turin, Italy
\item \Adef{orgII}M.V. Lomonosov Moscow State University, D.V. Skobeltsyn Institute of Nuclear, Physics, Moscow, Russia
\item \Adef{orgIII}Department of Applied Physics, Aligarh Muslim University, Aligarh, India
\item \Adef{orgIV}Institute of Theoretical Physics, University of Wroclaw, Poland
\end{Authlist}

\section*{Collaboration Institutes}
\renewcommand\theenumi{\arabic{enumi}~}
\begin{Authlist}
\item \Idef{org1}A.I. Alikhanyan National Science Laboratory (Yerevan Physics Institute) Foundation, Yerevan, Armenia
\item \Idef{org2}Bogolyubov Institute for Theoretical Physics, National Academy of Sciences of Ukraine, Kiev, Ukraine
\item \Idef{org3}Bose Institute, Department of Physics  and Centre for Astroparticle Physics and Space Science (CAPSS), Kolkata, India
\item \Idef{org4}Budker Institute for Nuclear Physics, Novosibirsk, Russia
\item \Idef{org5}California Polytechnic State University, San Luis Obispo, California, United States
\item \Idef{org6}Central China Normal University, Wuhan, China
\item \Idef{org7}Centre de Calcul de l'IN2P3, Villeurbanne, Lyon, France
\item \Idef{org8}Centro de Aplicaciones Tecnol\'{o}gicas y Desarrollo Nuclear (CEADEN), Havana, Cuba
\item \Idef{org9}Centro de Investigaci\'{o}n y de Estudios Avanzados (CINVESTAV), Mexico City and M\'{e}rida, Mexico
\item \Idef{org10}Centro Fermi - Museo Storico della Fisica e Centro Studi e Ricerche ``Enrico Fermi', Rome, Italy
\item \Idef{org11}Chicago State University, Chicago, Illinois, United States
\item \Idef{org12}China Institute of Atomic Energy, Beijing, China
\item \Idef{org13}Chonbuk National University, Jeonju, Republic of Korea
\item \Idef{org14}Comenius University Bratislava, Faculty of Mathematics, Physics and Informatics, Bratislava, Slovakia
\item \Idef{org15}COMSATS University Islamabad, Islamabad, Pakistan
\item \Idef{org16}Creighton University, Omaha, Nebraska, United States
\item \Idef{org17}Department of Physics, Aligarh Muslim University, Aligarh, India
\item \Idef{org18}Department of Physics, Pusan National University, Pusan, Republic of Korea
\item \Idef{org19}Department of Physics, Sejong University, Seoul, Republic of Korea
\item \Idef{org20}Department of Physics, University of California, Berkeley, California, United States
\item \Idef{org21}Department of Physics, University of Oslo, Oslo, Norway
\item \Idef{org22}Department of Physics and Technology, University of Bergen, Bergen, Norway
\item \Idef{org23}Dipartimento di Fisica dell'Universit\`{a} 'La Sapienza' and Sezione INFN, Rome, Italy
\item \Idef{org24}Dipartimento di Fisica dell'Universit\`{a} and Sezione INFN, Cagliari, Italy
\item \Idef{org25}Dipartimento di Fisica dell'Universit\`{a} and Sezione INFN, Trieste, Italy
\item \Idef{org26}Dipartimento di Fisica dell'Universit\`{a} and Sezione INFN, Turin, Italy
\item \Idef{org27}Dipartimento di Fisica e Astronomia dell'Universit\`{a} and Sezione INFN, Bologna, Italy
\item \Idef{org28}Dipartimento di Fisica e Astronomia dell'Universit\`{a} and Sezione INFN, Catania, Italy
\item \Idef{org29}Dipartimento di Fisica e Astronomia dell'Universit\`{a} and Sezione INFN, Padova, Italy
\item \Idef{org30}Dipartimento di Fisica `E.R.~Caianiello' dell'Universit\`{a} and Gruppo Collegato INFN, Salerno, Italy
\item \Idef{org31}Dipartimento DISAT del Politecnico and Sezione INFN, Turin, Italy
\item \Idef{org32}Dipartimento di Scienze e Innovazione Tecnologica dell'Universit\`{a} del Piemonte Orientale and INFN Sezione di Torino, Alessandria, Italy
\item \Idef{org33}Dipartimento Interateneo di Fisica `M.~Merlin' and Sezione INFN, Bari, Italy
\item \Idef{org34}European Organization for Nuclear Research (CERN), Geneva, Switzerland
\item \Idef{org35}Faculty of Electrical Engineering, Mechanical Engineering and Naval Architecture, University of Split, Split, Croatia
\item \Idef{org36}Faculty of Engineering and Science, Western Norway University of Applied Sciences, Bergen, Norway
\item \Idef{org37}Faculty of Nuclear Sciences and Physical Engineering, Czech Technical University in Prague, Prague, Czech Republic
\item \Idef{org38}Faculty of Science, P.J.~\v{S}af\'{a}rik University, Ko\v{s}ice, Slovakia
\item \Idef{org39}Frankfurt Institute for Advanced Studies, Johann Wolfgang Goethe-Universit\"{a}t Frankfurt, Frankfurt, Germany
\item \Idef{org40}Gangneung-Wonju National University, Gangneung, Republic of Korea
\item \Idef{org41}Gauhati University, Department of Physics, Guwahati, India
\item \Idef{org42}Helmholtz-Institut f\"{u}r Strahlen- und Kernphysik, Rheinische Friedrich-Wilhelms-Universit\"{a}t Bonn, Bonn, Germany
\item \Idef{org43}Helsinki Institute of Physics (HIP), Helsinki, Finland
\item \Idef{org44}High Energy Physics Group,  Universidad Aut\'{o}noma de Puebla, Puebla, Mexico
\item \Idef{org45}Hiroshima University, Hiroshima, Japan
\item \Idef{org46}Hochschule Worms, Zentrum  f\"{u}r Technologietransfer und Telekommunikation (ZTT), Worms, Germany
\item \Idef{org47}Horia Hulubei National Institute of Physics and Nuclear Engineering, Bucharest, Romania
\item \Idef{org48}Indian Institute of Technology Bombay (IIT), Mumbai, India
\item \Idef{org49}Indian Institute of Technology Indore, Indore, India
\item \Idef{org50}Indonesian Institute of Sciences, Jakarta, Indonesia
\item \Idef{org51}INFN, Laboratori Nazionali di Frascati, Frascati, Italy
\item \Idef{org52}INFN, Sezione di Bari, Bari, Italy
\item \Idef{org53}INFN, Sezione di Bologna, Bologna, Italy
\item \Idef{org54}INFN, Sezione di Cagliari, Cagliari, Italy
\item \Idef{org55}INFN, Sezione di Catania, Catania, Italy
\item \Idef{org56}INFN, Sezione di Padova, Padova, Italy
\item \Idef{org57}INFN, Sezione di Roma, Rome, Italy
\item \Idef{org58}INFN, Sezione di Torino, Turin, Italy
\item \Idef{org59}INFN, Sezione di Trieste, Trieste, Italy
\item \Idef{org60}Inha University, Incheon, Republic of Korea
\item \Idef{org61}Institut de Physique Nucl\'{e}aire d'Orsay (IPNO), Institut National de Physique Nucl\'{e}aire et de Physique des Particules (IN2P3/CNRS), Universit\'{e} de Paris-Sud, Universit\'{e} Paris-Saclay, Orsay, France
\item \Idef{org62}Institute for Nuclear Research, Academy of Sciences, Moscow, Russia
\item \Idef{org63}Institute for Subatomic Physics, Utrecht University/Nikhef, Utrecht, Netherlands
\item \Idef{org64}Institute for Theoretical and Experimental Physics, Moscow, Russia
\item \Idef{org65}Institute of Experimental Physics, Slovak Academy of Sciences, Ko\v{s}ice, Slovakia
\item \Idef{org66}Institute of Physics, Homi Bhabha National Institute, Bhubaneswar, India
\item \Idef{org67}Institute of Physics of the Czech Academy of Sciences, Prague, Czech Republic
\item \Idef{org68}Institute of Space Science (ISS), Bucharest, Romania
\item \Idef{org69}Institut f\"{u}r Kernphysik, Johann Wolfgang Goethe-Universit\"{a}t Frankfurt, Frankfurt, Germany
\item \Idef{org70}Instituto de Ciencias Nucleares, Universidad Nacional Aut\'{o}noma de M\'{e}xico, Mexico City, Mexico
\item \Idef{org71}Instituto de F\'{i}sica, Universidade Federal do Rio Grande do Sul (UFRGS), Porto Alegre, Brazil
\item \Idef{org72}Instituto de F\'{\i}sica, Universidad Nacional Aut\'{o}noma de M\'{e}xico, Mexico City, Mexico
\item \Idef{org73}iThemba LABS, National Research Foundation, Somerset West, South Africa
\item \Idef{org74}Johann-Wolfgang-Goethe Universit\"{a}t Frankfurt Institut f\"{u}r Informatik, Fachbereich Informatik und Mathematik, Frankfurt, Germany
\item \Idef{org75}Joint Institute for Nuclear Research (JINR), Dubna, Russia
\item \Idef{org76}Korea Institute of Science and Technology Information, Daejeon, Republic of Korea
\item \Idef{org77}KTO Karatay University, Konya, Turkey
\item \Idef{org78}Laboratoire de Physique Subatomique et de Cosmologie, Universit\'{e} Grenoble-Alpes, CNRS-IN2P3, Grenoble, France
\item \Idef{org79}Lawrence Berkeley National Laboratory, Berkeley, California, United States
\item \Idef{org80}Lund University Department of Physics, Division of Particle Physics, Lund, Sweden
\item \Idef{org81}Nagasaki Institute of Applied Science, Nagasaki, Japan
\item \Idef{org82}Nara Women{'}s University (NWU), Nara, Japan
\item \Idef{org83}National and Kapodistrian University of Athens, School of Science, Department of Physics , Athens, Greece
\item \Idef{org84}National Centre for Nuclear Research, Warsaw, Poland
\item \Idef{org85}National Institute of Science Education and Research, Homi Bhabha National Institute, Jatni, India
\item \Idef{org86}National Nuclear Research Center, Baku, Azerbaijan
\item \Idef{org87}National Research Centre Kurchatov Institute, Moscow, Russia
\item \Idef{org88}Niels Bohr Institute, University of Copenhagen, Copenhagen, Denmark
\item \Idef{org89}Nikhef, National institute for subatomic physics, Amsterdam, Netherlands
\item \Idef{org90}NRC Kurchatov Institute IHEP, Protvino, Russia
\item \Idef{org91}NRNU Moscow Engineering Physics Institute, Moscow, Russia
\item \Idef{org92}Nuclear Physics Group, STFC Daresbury Laboratory, Daresbury, United Kingdom
\item \Idef{org93}Nuclear Physics Institute of the Czech Academy of Sciences, \v{R}e\v{z} u Prahy, Czech Republic
\item \Idef{org94}Oak Ridge National Laboratory, Oak Ridge, Tennessee, United States
\item \Idef{org95}Ohio State University, Columbus, Ohio, United States
\item \Idef{org96}Petersburg Nuclear Physics Institute, Gatchina, Russia
\item \Idef{org97}Physics department, Faculty of science, University of Zagreb, Zagreb, Croatia
\item \Idef{org98}Physics Department, Panjab University, Chandigarh, India
\item \Idef{org99}Physics Department, University of Jammu, Jammu, India
\item \Idef{org100}Physics Department, University of Rajasthan, Jaipur, India
\item \Idef{org101}Physikalisches Institut, Eberhard-Karls-Universit\"{a}t T\"{u}bingen, T\"{u}bingen, Germany
\item \Idef{org102}Physikalisches Institut, Ruprecht-Karls-Universit\"{a}t Heidelberg, Heidelberg, Germany
\item \Idef{org103}Physik Department, Technische Universit\"{a}t M\"{u}nchen, Munich, Germany
\item \Idef{org104}Politecnico di Bari, Bari, Italy
\item \Idef{org105}Research Division and ExtreMe Matter Institute EMMI, GSI Helmholtzzentrum f\"ur Schwerionenforschung GmbH, Darmstadt, Germany
\item \Idef{org106}Rudjer Bo\v{s}kovi\'{c} Institute, Zagreb, Croatia
\item \Idef{org107}Russian Federal Nuclear Center (VNIIEF), Sarov, Russia
\item \Idef{org108}Saha Institute of Nuclear Physics, Homi Bhabha National Institute, Kolkata, India
\item \Idef{org109}School of Physics and Astronomy, University of Birmingham, Birmingham, United Kingdom
\item \Idef{org110}Secci\'{o}n F\'{\i}sica, Departamento de Ciencias, Pontificia Universidad Cat\'{o}lica del Per\'{u}, Lima, Peru
\item \Idef{org111}Shanghai Institute of Applied Physics, Shanghai, China
\item \Idef{org112}St. Petersburg State University, St. Petersburg, Russia
\item \Idef{org113}Stefan Meyer Institut f\"{u}r Subatomare Physik (SMI), Vienna, Austria
\item \Idef{org114}SUBATECH, IMT Atlantique, Universit\'{e} de Nantes, CNRS-IN2P3, Nantes, France
\item \Idef{org115}Suranaree University of Technology, Nakhon Ratchasima, Thailand
\item \Idef{org116}Technical University of Ko\v{s}ice, Ko\v{s}ice, Slovakia
\item \Idef{org117}Technische Universit\"{a}t M\"{u}nchen, Excellence Cluster 'Universe', Munich, Germany
\item \Idef{org118}The Henryk Niewodniczanski Institute of Nuclear Physics, Polish Academy of Sciences, Cracow, Poland
\item \Idef{org119}The University of Texas at Austin, Austin, Texas, United States
\item \Idef{org120}Universidad Aut\'{o}noma de Sinaloa, Culiac\'{a}n, Mexico
\item \Idef{org121}Universidade de S\~{a}o Paulo (USP), S\~{a}o Paulo, Brazil
\item \Idef{org122}Universidade Estadual de Campinas (UNICAMP), Campinas, Brazil
\item \Idef{org123}Universidade Federal do ABC, Santo Andre, Brazil
\item \Idef{org124}University College of Southeast Norway, Tonsberg, Norway
\item \Idef{org125}University of Cape Town, Cape Town, South Africa
\item \Idef{org126}University of Houston, Houston, Texas, United States
\item \Idef{org127}University of Jyv\"{a}skyl\"{a}, Jyv\"{a}skyl\"{a}, Finland
\item \Idef{org128}University of Liverpool, Liverpool, United Kingdom
\item \Idef{org129}University of Science and Techonology of China, Hefei, China
\item \Idef{org130}University of Tennessee, Knoxville, Tennessee, United States
\item \Idef{org131}University of the Witwatersrand, Johannesburg, South Africa
\item \Idef{org132}University of Tokyo, Tokyo, Japan
\item \Idef{org133}University of Tsukuba, Tsukuba, Japan
\item \Idef{org134}Universit\'{e} Clermont Auvergne, CNRS/IN2P3, LPC, Clermont-Ferrand, France
\item \Idef{org135}Universit\'{e} de Lyon, Universit\'{e} Lyon 1, CNRS/IN2P3, IPN-Lyon, Villeurbanne, Lyon, France
\item \Idef{org136}Universit\'{e} de Strasbourg, CNRS, IPHC UMR 7178, F-67000 Strasbourg, France, Strasbourg, France
\item \Idef{org137}Universit\'{e} Paris-Saclay Centre d'Etudes de Saclay (CEA), IRFU, D\'{e}partment de Physique Nucl\'{e}aire (DPhN), Saclay, France
\item \Idef{org138}Universit\`{a} degli Studi di Foggia, Foggia, Italy
\item \Idef{org139}Universit\`{a} degli Studi di Pavia, Pavia, Italy
\item \Idef{org140}Universit\`{a} di Brescia, Brescia, Italy
\item \Idef{org141}Variable Energy Cyclotron Centre, Homi Bhabha National Institute, Kolkata, India
\item \Idef{org142}Warsaw University of Technology, Warsaw, Poland
\item \Idef{org143}Wayne State University, Detroit, Michigan, United States
\item \Idef{org144}Westf\"{a}lische Wilhelms-Universit\"{a}t M\"{u}nster, Institut f\"{u}r Kernphysik, M\"{u}nster, Germany
\item \Idef{org145}Wigner Research Centre for Physics, Hungarian Academy of Sciences, Budapest, Hungary
\item \Idef{org146}Yale University, New Haven, Connecticut, United States
\item \Idef{org147}Yonsei University, Seoul, Republic of Korea
\end{Authlist}
\endgroup
\end{document}